\documentclass[Universe,conferencereport,accept,moreauthors,pdftex,10pt,a4paper]{mdpi}

\usepackage{booktabs} 
\usepackage{multirow}
\usepackage{soul} 
\usepackage{microtype}

\setitemize{parsep=6pt,itemsep=0pt,leftmargin=*,labelsep=5.5mm,align=parleft}
\setenumerate{parsep=6pt,itemsep=0pt,leftmargin=*,labelsep=5.5mm,align=parleft}

\usepackage{array} \newcommand{\PreserveBackslash}[1]{\let\temp=\\#1\let\\=\temp}
\newcolumntype{C}[1]{>{\PreserveBackslash\centering}m{#1}}
\newcolumntype{R}[1]{>{\PreserveBackslash\raggedleft}m{#1}}
\newcolumntype{L}[1]{>{\PreserveBackslash\raggedright}m{#1}}

\firstpage{1} 
\articlenumber{32}
\doinum{10.3390/universe3020032}
\pubvolume{3}
\pubyear{2017}
\copyrightyear{2017}
\externaleditor{{Academic Editors: Mariusz P. Dąbrowski, Manuel Krämer and Vincenzo Salzano}}
\history{{Received: 31 January 2017; Accepted: 17 March 2017; Published: 30 March 2017}}
\pdfoutput=1

 \theoremstyle{mdpi}
 \newcounter{thm}
 \newcounter{ex}
 \newcounter{re}
\Title{Probing the Gravitational Dependence of the Fine-Structure Constant from Observations of White~Dwarf Stars}

\Author{Matthew~B.~Bainbridge $^{1,}$*, Martin~A.~Barstow $^{1}$, Nicole~Reindl $^{1}$, {W.-Ü~Lydia~Tchang-Brillet} $^{2,3}$, Thomas~R.~Ayres $^{4}$, John~K.~Webb $^{5}$, John~D.~Barrow $^{6}$, Jiting~Hu $^{5}$, Jay B.~Holberg $^{7}$, Simon~P.~Preval $^{8}$, Wim~Ubachs $^{9}$, {Vladimir}~A.~Dzuba $^{5}$, {Victor}~V.~Flambaum $^{5}$, Vincent~Dumont~$^{10}$~and Julian~C.~Berengut $^{5}$}
\AuthorNames{Matthew~B.~Bainbridge, Martin~A.~Barstow,  Nicole~Reindl,  {W.-Ü~Lydia~Tchang-Brillet}$,  Thomas~R.~Ayres, John ~K.~Webb,  John~D.~Barrow, Jiting~Hu, Jay B.~Holberg, Simon~P.~Preval, Wim~Ubachs, {Vladimir}~A.~Dzuba, {Victor}~V.~Flambaum, Vincent~Dumont~and Julian~C.~Berengut}

\address{%
$^{1}$ \quad Department of Physics and Astronomy, University of Leicester, University Road, Leicester LE1 7RH, UK; mab@le.ac.uk (M.A.B.); nr152@le.ac.uk (N.R.)\\
$^{2}$ \quad LERMA, Observatoire de Paris-Meudon, PSL Research University, CNRS, UMR8112, F-92195 Meudon, France; lydia.tchang-brillet@obspm.fr\\
$^{3}$ \quad {Sorbonne Universités}, UPMC University Paris 6, UMR8112, LERMA, F-75005 Paris, France\\ 
$^{4}$ \quad Center for Astrophysics and Space Astronomy, University of Colorado, Boulder, CO 80309-0389, USA; Thomas.Ayres@Colorado.edu\\
$^{5}$ \quad School of Physics, University of New South Wales, Sydney, NSW 2052, Australia; jkw@phys.unsw.edu.au~(J.K.W.); jiting.hu@student.unsw.edu.au (J.H.); v.dzuba@unsw.edu.au (V.A.D.); v.flambaum@unsw.edu.au (V.V.F.); julianberengut@gmail.com (J.C.B.)\\
$^{6}$ \quad DAMTP, Centre for Mathematical Sciences, University of Cambridge, Wilberforce Road, \mbox{Cambridge CB3 0WA}, UK; j.d.barrow@damtp.cam.ac.uk\\
$^{7}$ \quad Lunar and Planetary Laboratory, Sonett Space Sciences Building, University of Arizona, Tucson, AZ 85721, USA; holberg@argus.lpl.arizona.edu\\
$^{8}$ \quad Department of Physics, University of Strathclyde, Glasgow G4 0NG, UK; simon.preval@strath.ac.uk\\
$^{9}$ \quad Department of Physics and Astronomy, LaserLab, VU University, De Boelelaan 1081, NL-1081~HV~Amsterdam, The Netherlands; w.m.g.ubachs@vu.nl\\
$^{10}$ \quad Department of Physics, University of California, Berkeley, CA 94720-7300, USA; vincentdumont11@gmail.com\\
}

\corres{Correspondence: mbb8@le.ac.uk}

\abstract{
Hot white dwarf stars 
are the ideal probe for a relationship between the fine-structure constant 
and strong gravitational fields, providing us with an opportunity for a 
direct observational test. We study a sample of hot white dwarf stars, 
combining far-UV spectroscopic observations, atomic physics, atmospheric 
modelling, and fundamental physics in the search for variation in the fine 
structure constant. This variation manifests as shifts in the observed 
wavelengths of absorption lines, such as quadruply ionized iron (FeV) and quadruply ionized nickel (NiV), when compared to 
laboratory wavelengths. {Berengut et al. (\emph{Phys. Rev. Lett.} \textbf{2013}, \emph{111}, 010801)}
~demonstrated the validity 
of such an analysis using high-resolution Hubble Space Telescope (HST)/Space Telescope Imaging Spectrograph (STIS) spectra of G191-B2B. We 
have made three important improvements by: (a) using three new independent 
sets of laboratory wavelengths; (b)~analysing a sample of objects; and 
(c) improving the methodology by incorporating robust techniques from 
previous studies towards quasars (the Many Multiplet method). A successful 
detection would be the first direct measurement of a 
gravitational field effect on a bare constant of nature. Here we describe 
our approach and present preliminary results from nine objects using both 
FeV and NiV. 
}


\begin{document}




A common feature of many schemes to unify the strong, electro-weak,
and gravitational forces of nature is the prediction of violation 
of local Lorentz invariance and the Einstein equivalence principle 
at high energy \citep{1989PhRvD..39..683K}. This can manifest 
itself as variations in the fundamental constants of physics 
(Newton's constant, $G$; proton-to-electron mass ratio, $\mu$; 
fine structure constant, $\alpha$; etc.) due to light scalar fields, 
the presence of extra space dimensions, or the non-uniqueness of the 
quantum vacuum state for the universe. Probing the variation of 
fundamental constants in the distant universe is an~important test 
of the equivalence principle and prospective theories of Grand Unification.

In a light scalar field the total mass and the total scalar charge are both 
proportional to the number of nucleons, for objects that are not too 
relativistic, so observing fundamental constants near gravitating massive 
bodies is one way to probe the form of a potential variation and the 
existence of scalar fields. However, the effect of a light scalar field on 
fundamental constants near massive bodies depends heavily on the theory 
being considered, particularly the type of coupling between the scalar 
fields and other fields \citep{2002PhLB..549..284M}. 
Flambaum and Shuryak (2008) \citep{2008AIPC..995....1F} considered a linear 
coupling between alpha and gravitational potential through the introduction 
of a massless scalar field, leading to the simple~relationship
\begin{displaymath}
\Delta \alpha / \alpha \equiv \frac{\alpha(r) - \alpha_0}{\alpha_0} \equiv k_
{\alpha} \Delta \phi = k_{\alpha} \Delta \left( \frac{GM}{rc^2} \right)
\end{displaymath}
where $\phi$ is the dimensionless gravitational potential ($\phi = \frac{GM}{rc^2}$), 
$k_{\alpha}$ is a dimensionless dependency parameter, $M$ is the 
mass of the object, $r$ is the radial distance from the object's center, and 
$\alpha_0$ is the laboratory value of the fine structure constant. 
If the relationship is indeed linear (or close to), then $k_{\alpha}$ is a 
constant and can be very accurately determined by high-precision atomic 
clocks \citep{2008AIPC..995....1F,2002PhRvD..65h1101B,2007PhRvA..76f2104F,
2007PhRvL..98g0801F,2008PhRvL.100n0801B,2008PhRvD..78f7304B,2012PhRvL.109h0801G,
2013PhRvL.111f0801L,2016arXiv160806050D}. 
However, $k_{\alpha}$ may not be constant \citep{2016PhRvD..94f4038M}, and the 
assumption that this relationship is linear needs testing. 
To probe a non-linear relationship, we need to observe $k_{\alpha}$ under conditions different 
 than those on Earth.

Hot white dwarf stars are the ideal probe for a relationship between 
$\alpha$ and strong gravitational fields. Hot white dwarfs---with masses 
comparable to the sun and radii comparable to Earth---generate strong 
gravitational fields and are typically bright (enough for precision 
spectroscopic analysis) with numerous absorption 
lines. Within the absorption spectra of white dwarfs, variation in $\alpha$ 
is manifested as shifts in the observed wavelengths of absorption lines when 
compared to laboratory wavelengths~\citep{1999PhRvL..82..888D}, providing 
us with an opportunity for a direct observational test. 

Berengut et al. (2013) \citep{2013PhRvL.111a0801B} recently used Hubble Space Telescope (HST)/Space Telescope Imaging Spectrograph (STIS) 
spectra of the hot white dwarf star G191-B2B to constrain 
$\Delta \alpha/\alpha$, by observing the wavelength shifts in 96 quadruply ionized iron (FeV) and 32 
quadruply ionized nickel (NiV) absorption features and deriving a separate limit for each metal: 
$\Delta \alpha/\alpha = (4.8 \pm 1.6) \times 10^{-5}$ for FeV and 
$\Delta \alpha/\alpha = (-6.1 \pm 5.8) \times 10^{-5}$ for NiV. 
Berengut et al. (2013) \citep{2013PhRvL.111a0801B} suggest that this 
inconsistency is due to a systematic effect in the laboratory wavelengths used. 
We have extended this work by: (a)~using new laboratory wavelengths; 
(b) analysing a sample of objects rather than a single object; and 
(c) refining the analysis methodology by incorporating robust techniques 
from previous studies towards quasars (the Many Multiplet method 
\citep{1999PhRvL..82..884W,2001PhRvL..87i1301W,2001MNRAS.327.1208M}).

We are using three new independent lists of laboratory wavelengths to 
investigate the suspected systematic gain calibration error suspected by 
Berengut et al. (2013) \citep{2013PhRvL.111a0801B}. This apparent systematic 
effect is an 
important problem, because the effect is dependent on transition wavelength---just like a~wavelength shift due to $\Delta \alpha/\alpha$. 
In 2013, there were two lists of laboratory wavelengths available for 
FeV and NiV with reasonable precision and within the wavelength range of 
interest: Ekberg (1975)~\citep{1975PhyS...12...42E} for FeV and 
Raassen \& van Kleff \citep{1976PhyBC..85..180R} for NiV. 
Since then, three new lists have become available: 
the~(a)~Kramida; (b) Tchang-Brillet; and (c) Nave wavelength lists. Kramida (2014) 
\citep{2014ApJS..212...11K} published an~updated list of laboratory 
wavelengths for FeV, based on more recent observations by Azarov et al. 
(2001) \citep{2001PhyS...63..438A} (outside of our wavelength range of 
interest) in addition to the Ekberg (1975) \citep{1975PhyS...12...42E} 
laboratory results. Between 2014--2015 Tchang-Brillet (LERMA, {Meudon},{ France}) 
and a group led by Gillian Nave (NIST, Gaithersburg, {MD, USA}) independently 
re-observed the FeV laboratory wavelengths 
\citep{2015AAS...22533903W,2015IAUGA..2253006W,2016AAS...22724402W}. 
The team at NIST also re-observed the NiV laboratory wavelengths. The 
apparent systematic effect in the Raassen NiV laboratory wavelengths noted by 
Berengut et al. (2013) \citep{2013PhRvL.111a0801B} does not appear to 
be present in the new Nave NiV wavelengths from NIST.

In order to study a broader compactness range and to enlarge the size of our 
sample, we conducted a search of both the literature and the Mikulski Archive 
for Space Telescopes (MAST). We used the~following selection criteria: (a) 
photospheric absorption lines of FeV or NiV (atomic transitions for which we 
have new accurate laboratory wavelengths); (b) observed in the far-UV (the 
wavelength range of the FeV and NiV absorption lines) using HST/STIS Echelle 
spectroscopy; and (c) signal-to-noise ratio greater than 30 (a threshold for 
reasonable statistical uncertainties we determined using numerical 
simulations). We found that only the HST/STIS Echelle spectra provide the 
necessary wavelength accuracy (1 km$\cdotp$s$^{-1}$ \citep{2010ApJS..187..149A}) 
needed for this project. In addition to G191-B2B (the object studied in 
\citep{2013PhRvL.111a0801B}), we identified nine hot, bright white dwarfs 
and sub-dwarfs. We were also awarded 12 orbits with HST/STIS (scheduled for 
autumn 2017) to obtain new far-UV observations of three bright white dwarfs 
known to have photospheric Fe and Ni absorption lines. Table \ref{tab2} 
provides an overview of the~13~objects that will be studied in the course 
of this project. Our sample includes objects with gravitational potentials 
spanning four orders of magnitude.

\begin{table}[H]
\caption{
Characteristics of the white dwarf and sub dwarf sample. 
Uncertainties are 1$\sigma$. $g = GM/r^2$, the surface gravity in cm$\cdotp$s$^{-2}$. 
\label{tab2}}
\centering
{\scriptsize
\begin{tabular}{llrrlcl}
\toprule
\multirow{2}{*}{\bf \textbf{Object}     }          & \multirow{2}{*}{\bf \textbf{Type}}  & \textbf{RA (J2000)} &  \textbf{Dec. (J2000)} & \hspace{5.00mm} \boldmath \textbf{$T_{eff}$} & \multirow{2}{*}{\bf  \boldmath \textbf{$\log$} \textbf{g}} &\multirow{2}{*}{\bf \textbf{Ref.}}  \\
                    &      & \textbf{(Degrees)}\hspace{0.25mm}  & \textbf{(Degrees)}\hspace{0.75mm}    & \hspace{5.00mm} \textbf{(K)}      &          &      \\
\midrule	           
vz 1128             & O(H) & 205.569792 & 28.433639    & $36,600 \pm 400$   & $3.9 \pm 0.1$   & \cite{2015MNRAS.452.2292C} \\
ROB 162             & O(H) & 265.159792 & $-$53.642111   & $51,000 \pm 2000$  & $4.5 \pm 0.2$   & \cite{1986AA...169..244H} \\
BD + 28$^{\circ}$4211 & sdO  & 327.795813 & 28.863847    & $82,000 \pm 5000$  & $6.20 \pm 0.15$ & \cite{2013ApJ...773...84L} \\
Sh 2-174            & O(H) & 356.260417 & 80.950000    & $64,000 \pm 2900$  & $6.94 \pm 0.16$ & \cite{2005MNRAS.363..183G} \\
Sh2-313             & DAO  & 193.386496 & $-$22.872984   & $80,000 \pm 10,000$ & $7.2 \pm 0.3$   & \cite{2012AA...548A.109Z} \\
HS0505 + 0112         & DAO  & 77.128458  & 1.277611     & $63,200 \pm 2100$  & $7.30 \pm 0.15$ & \cite{2005MNRAS.363..183G} \\
Ton21               & DA   & 145.711333 & 26.016647    & $69,710 \pm 530$   & $7.47 \pm 0.05$ & \cite{2003MNRAS.341..870B} \\
Feige 24            & DA   & 38.781522  & 3.732415     & $60,000 \pm 1100$  & $7.50 \pm 0.06$ & \cite{2003MNRAS.341..870B} \\
G191-B2B            & DA   & 76.377645  & 52.831215    & $52,500 \pm 900$   & $7.53 \pm 0.09$ & \cite{2003MNRAS.341..870B} \\
REJ0558-373         & DA   & 89.560542  & $-$37.573561   & $59,500 \pm 2200$  & $7.70 \pm 0.14$ & \cite{2003MNRAS.344..562B} \\
RE-J0623-371 *       & DA   & 95.800417  & $-$37.691389   & $58,200 \pm 1800$  & $7.14 \pm 0.11$ & \cite{2003MNRAS.344..562B} \\
REJ2214-492 *        & DA   & 333.549642 & $-$49.324239   & $61,600 \pm 2300$  & $7.29 \pm 0.11$ & \cite{2003MNRAS.344..562B} \\
REJ0457-281 *        & DA   & 74.307917  & $-$28.131667   & $51,000 \pm 1100$  & $7.93 \pm 0.08$ & \cite{2003MNRAS.344..562B} \\
\bottomrule
\end{tabular}\\
}
\begin{tabular}{ccc}
\multicolumn{1}{c}{\footnotesize *    To be observed with Hubble Space Telescope (HST)/Space Telescope Imaging Spectrograph (STIS) during cycle 24.}
\end{tabular}
\end{table}

We examine the spectral data of each object before fitting the absorption 
lines to identify the FeV and NiV transitions relevant for estimation of 
$\Delta\alpha/\alpha$. All transitions were visually checked for obvious 
cases of blends, and where found, those transitions were discarded. In addition, 
for the purposes of the preliminary results in this paper, we confine 
ourselves to using the FeV and NiV transitions listed in Berengut et al.
(2013) \citep{2013PhRvL.111a0801B}. However, we use the three new laboratory 
wavelengths (the Kramida, Tchang-Brillet, and Nave wavelength lists) 
available for these transitions as discussed above.

We fit the absorption spectra in the normal way, using the Many Multiplet 
method and the software {\sc VPFIT} \footnote{R. F. Carswell and J. K. Webb, 2015, \url{http://www.ast.cam.ac.uk/~rfc/vpfit.html}.}
For each object, we initially construct a Voigt profile model (by visual 
inspection) with a single velocity component (absorption line), including 
all suitable transitions. We then apply {\sc VPFIT} to optimise the Voigt 
profile parameters, including $\Delta \alpha/\alpha$ as a free parameter 
in the fit. Statistical uncertainties are determined 
from the diagonal terms of the covariance matrix at the best-fitting~solution.

Our analysis is on-going, but  shows the preliminary results.
These preliminary results serve to highlight the importance of this kind 
of analysis. However, at this early stage, it would be premature to draw 
any conclusions about the relationship between $\Delta\alpha/\alpha$ and 
gravitational potential. We do not include a table of 
$\Delta \alpha/\alpha$ estimates or a weighted mean for this reason.

\begin{figure}[H]
\centering
\includegraphics[trim=0.15cm 1.00cm 5.30cm 0.15cm, clip=true, width=0.50\textwidth]{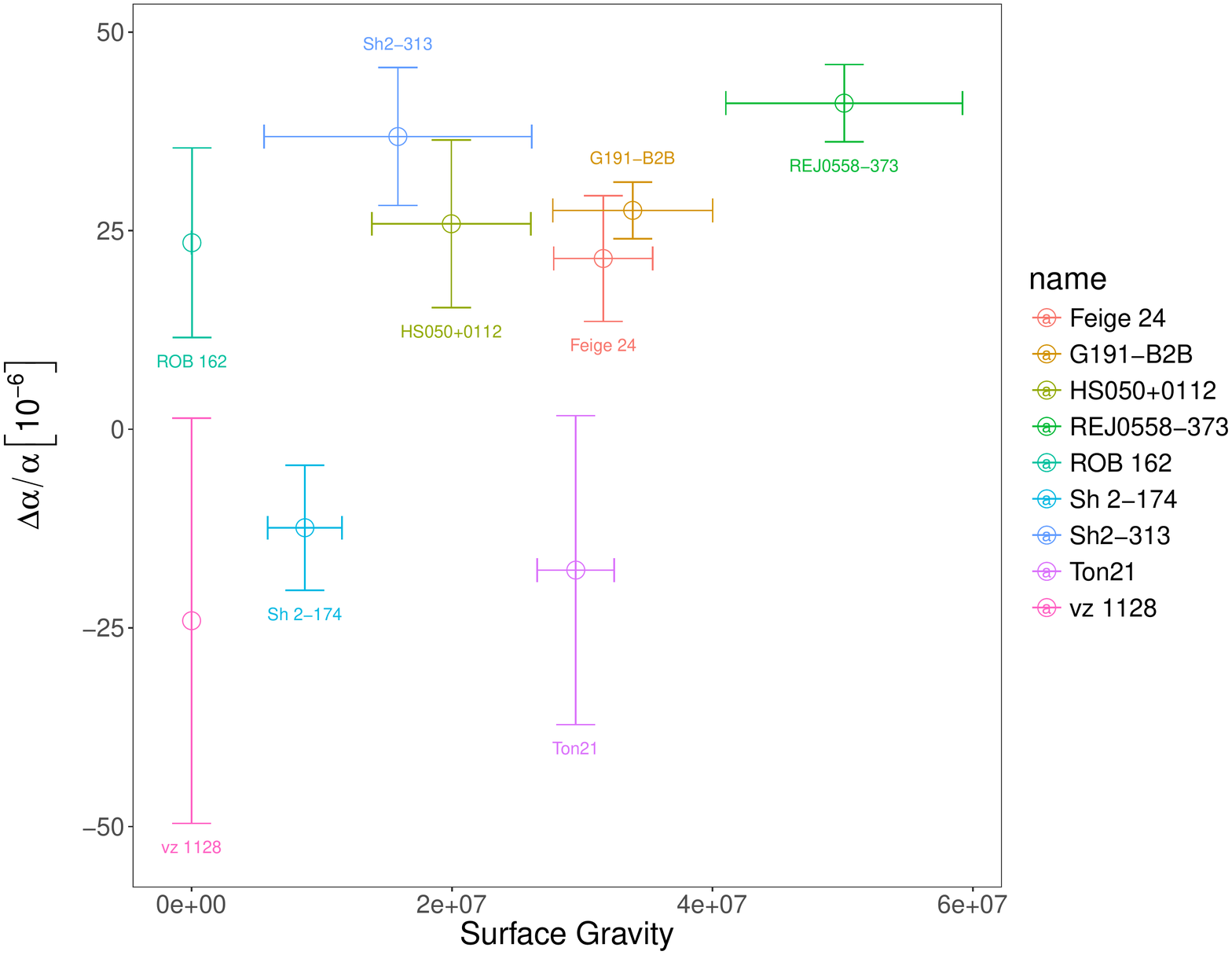}\\
\ \ \ \ \ \ \ \ g [cm s$^{-2}$]
\caption{
\label{fig1}
{\small
Preliminary $\Delta \alpha/\alpha$ results. Variation in the fine structure constant vs estimated surface gravity, \mbox{$g=GM/r^2$.} 
For each object in the sample we fitted Voigt profiles simultaneously to 
the relevant quadruply ionized iron (FeV) and quadruply ionized nickel (NiV) absorption lines of each object and estimated 
$\Delta \alpha/\alpha$ using {\sc VPFIT}. Here we show an example of our 
preliminary results, using the Ward and Nave (2015) \citep{2015AAS...22533903W} 
laboratory wavelengths. We see a similar trend using the Kramida and Tchang-Brillet 
wavelength lists. Error bars are 1 $\sigma$. The error bars on surface 
gravity for ROB 162 and vZ 1128 (both sub-dwarf objects) are too small to be 
seen {in this plot}. 
}
}
\end{figure}

A detailed consideration of possible systematic effects is required. The 
results shown in {Figure~\ref{fig1}} neglect several possible sources of systematic 
effects, which may explain this apparent detection of variation in $\alpha$: 
imprecise wavelength calibration, long and short range wavelength 
distortions, and~systematic effects in the laboratory wavelengths (despite 
the new measurements). It is important that these possible sources of bias 
are investigated. Additionally, these preliminary results include only 9 of 
the 13 objects in our sample. The remaining four objects (BD + 28$^{\circ}$4211, 
RE-J0623-371, REJ2214-492, and REJ0457-281) represent four of the five best targets 
for this analysis.

Studies such as the one summarised here provide a unique way of constraining 
new ideas in fundamental physics. The equivalence principle is at the heart 
of general relativity, and it fixes the fundamental constants of nature into 
an absolute unvarying structure---a structure independent of the material 
content of the universe. Probing the variation of fundamental constants tests 
the deepest depths of our current knowledge of physics, with the possibility 
of illuminating the next frontier of physics. This study is the first 
statistical sample of constraints on alpha from high-resolution white dwarf 
spectra.  Forthcoming HST observations will enhance that sample and improve 
the constraints~further.

\vspace{6pt} 

\acknowledgments{
We thank Gillian Nave (NIST) and Jacob Ward (NIST) for providing preliminary 
FeV and NiV wavelengths prior to publication. 
This research used the ALICE High Performance Computing Facility at the 
University of Leicester. 
This project is funded by a Leverhulme Trust Research Grant. 
WULTB wishes to acknowledge support from the LABEX Plas@par managed by the 
French ANR (ANR-11-IDEX-0004-02). 
{J.D.~Barrow is supported by the STFC of the UK.
}
}

\authorcontributions{JDB, MAB and JKW conceived the project. MAB leads and 
supervises the project. JB contributed to the concept and design of the
project. MBB and NR performed the data analysis and interpretation.
TRA contributed the co-added spectral data. WULTB contributed new FeV
laboratory wavelengths. JB, V. Dzuba and VF contributed to the
theoretical background and alpha sensitivity parameters for the
atomic transitions. V. Dzuba, VF and JKW invented the Many Multiplet
method used in this work and first demonstrated the advantages of
this method. JKW, JH, JBH, SPP, JB and V. Dumont provided discussion of
methodology and potential systematic effects. MBB wrote the paper. WU
provided critical revision of the paper. All authors commented on the
manuscript at all stages and approved the final version to be
published.}

\conflictsofinterest{The authors declare no conflict of interest.}

\abbreviations{%
\noindent The following abbreviations are used in this manuscript: \\
\noindent HST, Hubble Space Telescope, \\
\noindent STIS, Space Telescope Imaging Spectrograph, \\
\noindent LERMA, Laboratoire d’Etudes du Rayonnement et de la Matière en Astrophysique et Atmosphères, \\
\noindent NIST, National Institute of Standards and Technology, and \\
\noindent {\sc VPFIT}, Voigt Profile FITting software. \\
}

\bibliographystyle{mdpi}

\renewcommand\bibname{References}


\end{document}